\documentclass{iopart}

\usepackage{iopams}
\usepackage{graphicx}

\begin{document}

\title{Modelling DNA at the mesoscale: a challenge for nonlinear science?}

\author{Michel Peyrard$^{1}$,
Santiago Cuesta-L\'opez$^{1}$,
Guillaume James$^{2}$}

\address{$^{1}$ Universit\'e de Lyon, Ecole Normale Sup\'erieure de Lyon,
Laboratoire de Physique CNRS,
46 all\'ee d'Italie, 69364 Lyon Cedex 07, France
\ead{Michel.Peyrard@ens-lyon.fr}
}

\address{$^{2}$ Institut de Math\'ematiques de Toulouse, UMR 5219, D\'epartement de Math\'ematiques, INSA de Toulouse, 135 avenue de Rangueil, 31077 Toulouse Cedex 4, France}

\begin{abstract}
When it is viewed at the scale of a base pair, DNA appears as a nonlinear lattice. Modelling its properties is a fascinating goal. The detailed experiments that can be performed on this system impose constraints on the models and can be used as a guide to improve them. There are nevertheless many open problems, particularly to describe DNA at the scale of a few tens of base pairs, which is relevant for many biological phenomena.
\end{abstract}

\section{Introduction}  
\label{sec:intro}

\begin{figure}
  \centering
  \includegraphics[height=6.0cm]%
{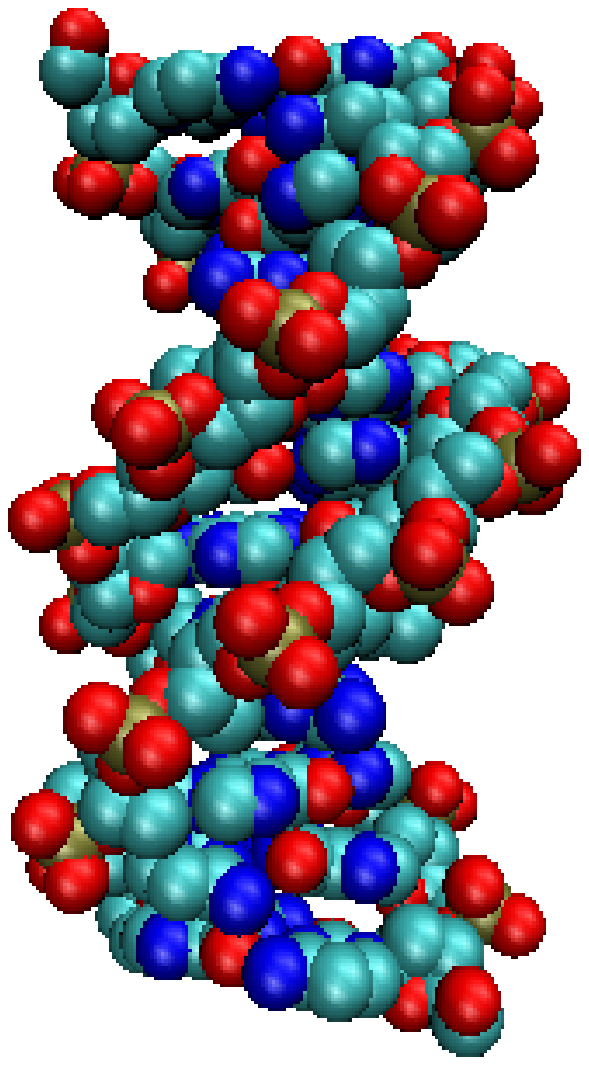} 
\hspace{1cm} \includegraphics[height=6cm]%
{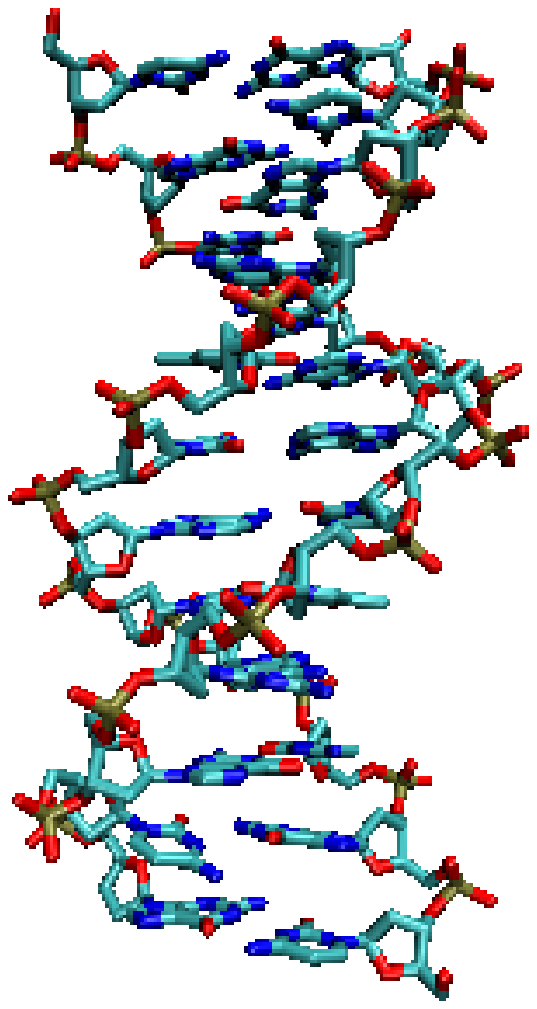}
  \caption{The structure of DNA in its B form
   in a full atomic representation (left)
    or in a schematic diagram showing the covalent interatomic bonds (right). The hydrogen bonds connecting the bases in pairs are not shown.}
\label{fig:struct}
\end{figure}

DNA is the molecule that encodes the information that
organisms need to live and reproduce themselves.
Fifty years after the discovery of its double helix structure
\cite{WATSONCRICK} it is still fascinating
physicists, as well as biologists, who try to unveil its
remarkable properties. But the {\em structure} of the molecule is only a static picture and it is now understood that the {\em dynamics} of biological molecules is also essential for their function. This is particularly true for DNA. The genetic code, defined by the sequence of bases pairs which form the plateaus of the double helix (Fig.~\ref{fig:struct})
is hidden in the core of the helix. When a gene is transcribed the hydrogen bonds that link the two bases in a base pair are broken in a region called the ``transcription bubble''and the bases are exposed outside of the helix for chemical reaction. This is possible because those bonds that hold the bases together are weak. As a result, 
even in the absence of enzymes involved in the
reading or duplication of the code, DNA undergoes large amplitude
fluctuations. The lifetime of a base pair, i.e.\ the time during which
it stays closed, is only of the order of a few milliseconds. 
Experiments show that
these fluctuations, known by biologists as the ``breathing of DNA'', are
highly localized, and may open a single base pair while the adjacent ones stay closed.

\medskip
Therefore, when it is viewed at the scale of a base pair, DNA appears as a rather regular lattice, which undergoes very large amplitude motions so that the nonlinear properties of the bonds that connect its elements cannot be ignored. Here we show that DNA is an attractive system for nonlinear science because its properties can be probed very accurately by experiments that combine the methods of physics with biological tools. The experimental results impose constraints on the theoretical modelling and therefore help us in the design of an appropriate model. Although significant progress have been made since the first attempts to describe DNA with nonlinear models \cite{ENGLANDER}, there are however many points that stay open and raise interesting questions for nonlinear physics.

\section{A simple nonlinear-lattice model of DNA}
\label{sec:model}

DNA has attracted the interest of physicists for a long time. One intriguing theoretical question is its ``melting'', i.e.\ the thermal separation of the two strands, which is a phase transition in a one-dimensional system \cite{POLAND66a,POLAND66b}. As long as one is only concerned by the open or closed state of a base pair, an Ising model in which the pair is a two-state system is sufficient. To study the nonlinear dynamics and fluctuations of the molecule, it is necessary to go beyond this oversimplified view. The next step is to keep a single degree of freedom per base pair, but use a real number, representing the stretching of the distance between the two bases instead of an integer taking only th values 0 or 1. This choice may seem extremely reductive and raises the question of what we mean by ``modelling'' a physical system, or, in other words the kind of question that we are asking.
This picture won't tell us anything about the actual motion of the bases in space, but it will nevertheless help us to understand the role of entropy versus energy in the melting of DNA, or the origin of the localisation that can lead to the large amplitude local fluctuations occuring in its ``breathing''. However, even if we select a very simple description of the molecule, the model cannot be chosen arbitrarily. As discussed below, the possibility to obtain detailed experimental data for DNA introduces constraints on the model.

\begin{figure}[h!]
  \centerline{
\includegraphics[width=6cm]{%
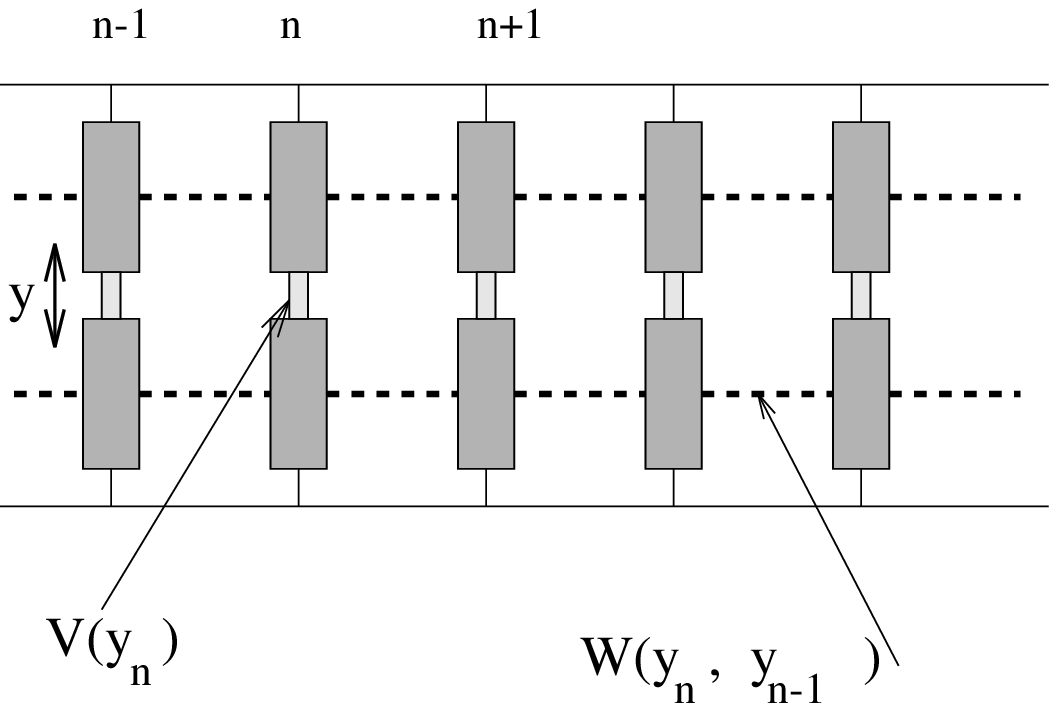}
\hspace{0.5cm}
\includegraphics[width=5.0cm]{%
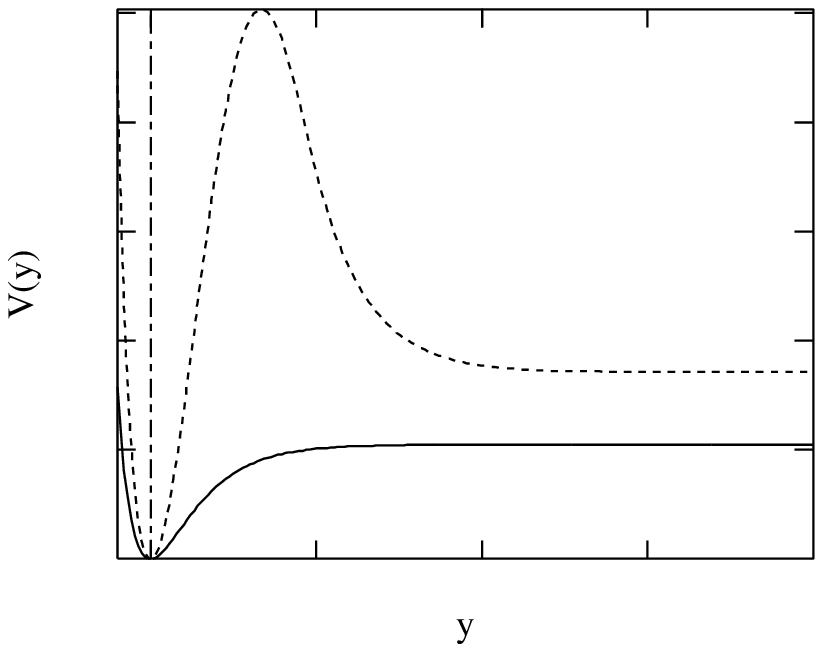}
}
\caption{The simple dynamical model for DNA nonlinear dynamics, described by
  Hamiltonian (\ref{eq:PBhamiltonian}). The right panel shows the Morse potential (full line) and the potential defined by Eq.~\ref{eq:potj} (dotted line). The parameters of the two potentials are chosen such that the models using these potentials give the same denaturation temperature. }
\label{fig:PBmodel}
\end{figure}

The Hamiltonian of the simplest model at the scale of
a base pair \cite{MPnonlinearity}, originally introduced to study the
thermal denaturation of DNA, is 
\begin{equation}
  \label{eq:PBhamiltonian}
H = \sum_n \frac{p_n^2}{2m} + W(y_n,y_{n-1}) + V(y_n), \quad{\mbox{
    with}}\quad p_n = m \frac{dy_n}{dt} \; ,
\end{equation}
where $n$ is the index of a base pair, $y_{n}$ its stretching, and 
$m$ its reduced mass. This model (Fig.~\ref{fig:PBmodel}) does not attempt to describe the actual geometry of the molecule and its Hamiltonian is simply chosen to include the main energetic contributions, 
the kinetic energy and the two dominant interaction energies.
The potential $V(y)$ describes the interaction between the two bases in a
pair. The model of DNA melting \cite{MPnonlinearity} uses a Morse potential
\begin{equation}
  \label{eq:MorsePot}
V(y) = D \left( e^{-\alpha y} - 1 \right)^2 \; ,
\end{equation}
where $D$ is the dissociation energy of the pair and $\alpha$ a parameter,
homogeneous to the inverse of a length, which sets the spatial scale of the
potential. This expression has been chosen because it is a standard expression
for chemical bonds and, moreover, it has the {\em appropriate qualitative
  shape}:
 (i) it includes a strong repulsive part for $y<0$, corresponding to the
  steric hindrance between bases in a pair,
 (ii) it has a minimum at the equilibrium position $y=0$,
 (iii) it becomes flat for large $y$, giving a force between the bases that
 tends to vanish, as expected when the bases are very far apart; this feature
 allows a complete dissociation of the base pair, which would be forbidden if
 we had chosen a simple harmonic potential.
The potential $W(y_n, y_{n-1})$ describes the interaction between adjacent
bases along the DNA molecule. It includes several contributions such as the overlap of the the $\pi$-electron orbitals of the organic rings that make up the bases and the coupling through the sugar-phosphate backbone of the DNA strands. The simplest model is obtained by assuming a harmonic coupling
\begin{equation}
\label{eq:harmW}
W(y_n, y_{n-1}) = \frac{1}{2} K (y_n - y_{n-1})^{2} \; .
\end{equation}
In its simplest form the model does not include the genetic code. All base pairs are considered to be
the same so that the results that we discuss in this section are only valid for homopolymers.

\begin{figure}[h]
\begin{center}
\begin{tabular}{cc}
\textbf{(a)} & \textbf{(b)} \\
\includegraphics[height=5cm]{%
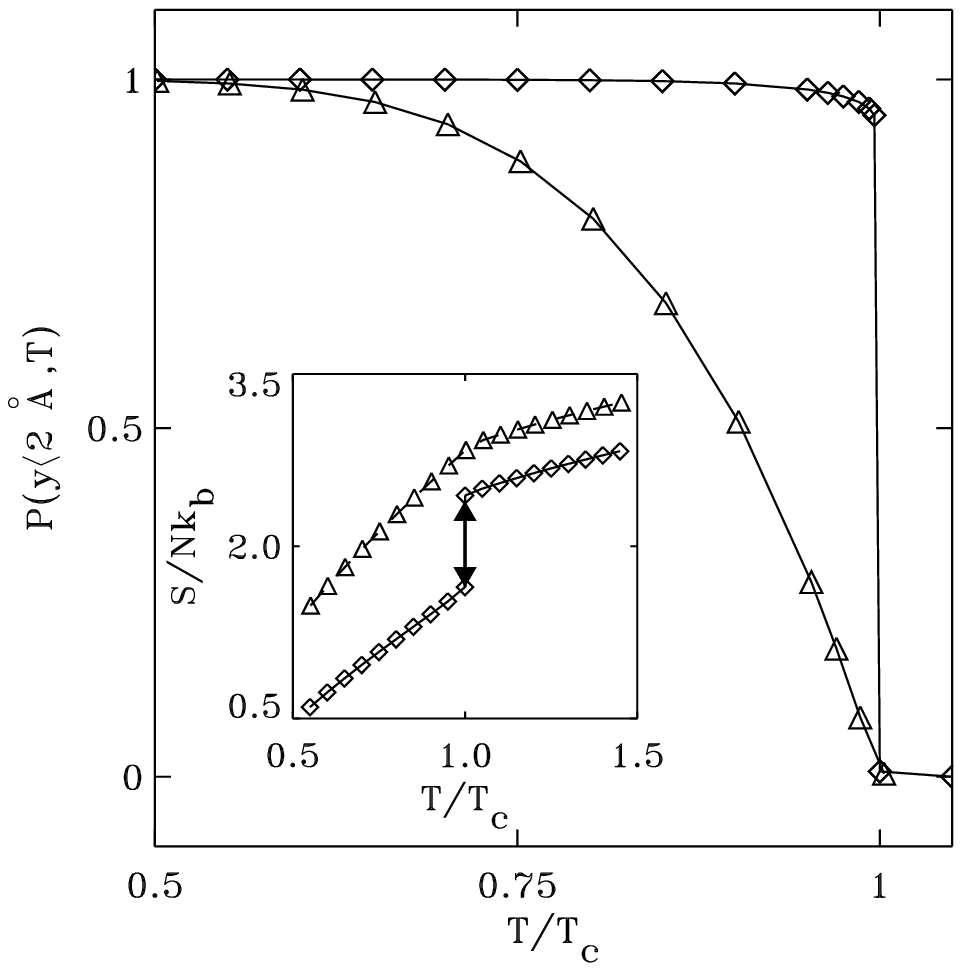} &
\includegraphics[height=5cm]{%
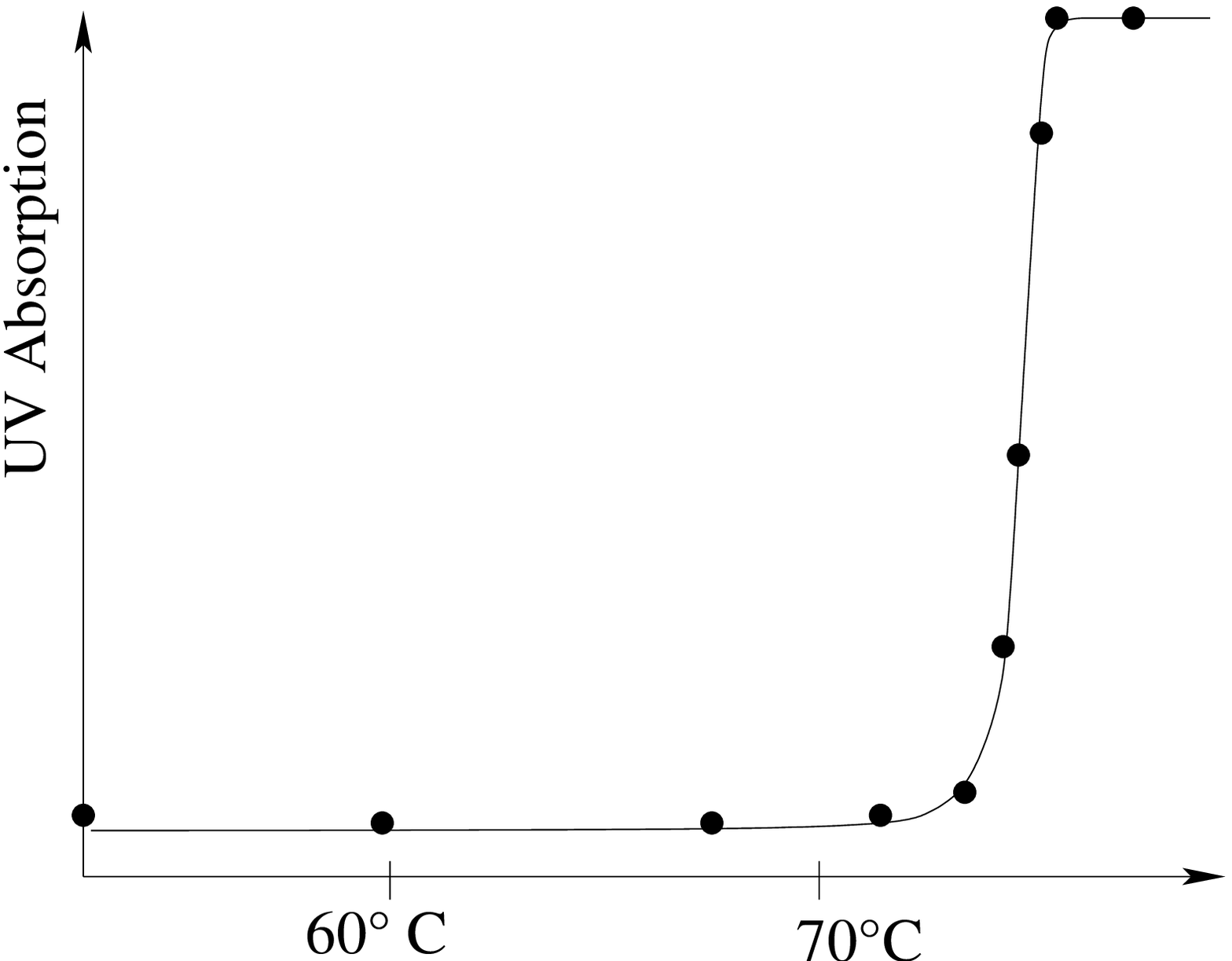}
\end{tabular}
\end{center}
\caption{a) Theoretical result for the fraction of closed base pairs versus
  temperature (probability $P(y<y_{0})$ ) for a model
  with harmonic coupling (triangles) or nonlinear coupling (diamonds). Inset:
  variation of the entropy versus temperature for the two models.
b) Denaturation curve of an homopolymer of DNA having only $G-C$ base
  pairs. (Figure adapted from Ref.~\cite{INMAN})
}
\label{fig:brokenBP}
\end{figure}

\medskip
The interest of this model is that it can be studied analytically or with simple numerical calculations \cite{MPnonlinearity,TPM}, and that it shows properties which are in qualitative agreement with the observations made on DNA. A statistical physics analysis shows that it does have a one-dimensional phase transition, corresponding to DNA melting because the mean value of the base pair stretching $\langle y \rangle$ diverges at a finite temperature $T_{c}$. The fraction of the base pairs that stay bound at a given temperature, defined as the base pairs having an average stretching below a threshold $y_{0}$, decreases smoothly from 1 at low temperature to 0 at $T_{c}$ as shown on Fig.~\ref{fig:brokenBP}. Moreover numerical simulations of the model in contact with a thermal bath show the presence of large amplitude, localised fluctuations of the stretching, which are strongly reminiscent of the experimentally observed ``breathing'' of DNA. These motions can be understood in terms of the localised modes, or discrete breathers, that we expect in such a lattice from nonlinear theory \cite{MACKAY94}.
Therefore it seems that a simple nonlinear model is exceptionally successful in reproducing the main observations associated to DNA fluctuations and thermal denaturation \cite{MADDOX89}. Moreover the analysis of the model gives us some insight on the origin of these observations, for instance by connecting breathing to nonlinear localisation.

\section{Experiments guide model improvements}
\label{seq:experimental}

{\em However the theoretician has to become more modest when he makes a quantitative comparison with experiments.} For DNA this test is demanding because experiments can provide numerous and precise data by combining methods from physics and biology.

First the thermal denaturation of the molecule can be easily monitored because, when the base pairs are broken, the UV absorption of a DNA solution increases sharply. Moreover, as DNA can be synthetised it is possible to study homopolymers, as described by the model above. Figure~\ref{fig:brokenBP}-b shows a typical result which indicates that the melting transition is very sharp. On the contrary the theoretical analysis finds a denaturation which extends over a broad temperature range $0.6 \, T_{c} \lesssim T \lesssim T_{c}$. Choosing appropriate model parameters can match the theoretical value of $T_{c}$ with the experimental melting temperature, but this is not sufficeint to ensure the validity of the simple model, which has to be corrected.

This can be done by improving the description of the stacking interaction. The harmonic expression (\ref{eq:harmW})
resulted from an expansion of the interaction potential around its
minimum. But, as previously mentioned, in DNA one base pair can be open while
its neighbors are not, which implies large values of the difference
$(y_n - y_{n-1})$. A small amplitude expansion is too rough.
To derive a more appropriate potential one has to think at its physical
origin. A large contribution to the stacking interaction comes from the
overlap  of the $\pi$ electrons of the base plateaus. When a base moves out of
the stack, the overlap decreases and its interaction with the neighboring
bases weakens. The redistribution of the $\pi$ electron which occurs when
hydrogen bonds are broken also contributes to this effect. This specificity
of the stacking interaction can be described by an improved potential
\begin{equation}
 \label{eq:nlstacking}
W(y_n, y_{n-1}) = \frac{1}{2} K \left( 1 + \rho e^{- \alpha (y_n + y_{n-1}) }
\right) (y_n - y_{n-1})^2 \;.
\end{equation}
The {\em plus} 
sign in the exponential term is important. As soon as either one of the two interacting base pairs is open (an
not necessarily both simultaneously) the effective coupling constant
drops from $K' \approx K(1 + \rho)$ down to $K' = K$. 
This fairly simple change in the model brings a very big quantitative improvement because the transition can become almost infinitely sharp if the parameters are properly chosen \cite{TDP} (Fig.~\ref{fig:brokenBP}). The role of the nonlinear stacking to modify the
character of the transition can be understood from simple qualitative
arguments. Above denaturation, the stacking gets weaker, reducing the rigidity
of the DNA strands, i.e.\ increasing their entropy. Although there is a
energetic cost to open the base pairs, if the entropy gain is large enough,
the opening may bring a gain in free energy $F = U - T S$, where $U$ is
the internal energy of the system and $S$ its entropy. The nonlinear stacking
leads to a kind of ``self-amplification process'' in the transition: when the
transition proceeds, the extra flexibility of the strands makes it easier,
which sustains the denaturation that becomes very sharp. This points out the importance of entropic effects in the thermal denaturation of DNA, as noticed also when a more macroscopic analysis of the transition is performed \cite{MUKAMEL}.

\medskip
But experiments can do much more than observing the average number of broken base pairs versus temperature.
The breathing of DNA can be studied accurately using proton-deuterium
exchange. If DNA is dissolved into deuterated water, when the opening
of a base pair exposes 
the protons that form the hydrogen bonds to the solvent, 
the so-called imino protons, those protons
can be exchanged with deuterium from the water molecules of the
solvent. The exchange rate can be accelerated by a catalyst. Then NMR
can be used to detect the deuterium atoms within the 
DNA molecule \cite{Leroy}. Kinetic  experiments
show that the lifetime of base pairs (time during which they stay
closed) is in the range of milliseconds at $35^{\circ}$C and 10 times
more at $0^{\circ}$C. The  lifetime of the open state is in the nanosecond
range (estimations are in the range 30 to 300 nanoseconds)
\cite{GUERON1990}. 
At biological temperatures the experiments show that {\em single base pair opening events are the only mode of base pair
  disruption.} This indicates that the large amplitude conformational
change that is able to break a pair and expose a base to the solvent is
a {\em highly localised phenomenon}. It should also be pointed out
that these localised fluctuations are observed in artificial molecules
which contain homogeneous sequences of identical base pairs. Therefore
the localisation cannot be attributed to the inhomogeneity of natural
DNA sequences. This supports the analysis suggested by the model, which attributes this localisation to nonlinearity.
However this result points out another major weakness the model.
Within this model, the opening appears as a
breathing mode so that a base pair opens and closes with a period of
the order of the slowest vibrational motions of DNA,
i.e.\ of the order of a picosecond, which is far below the observed time scales, although the breather itself can have a lifetime of
hundreds of picoseconds.

\begin{figure}[h]
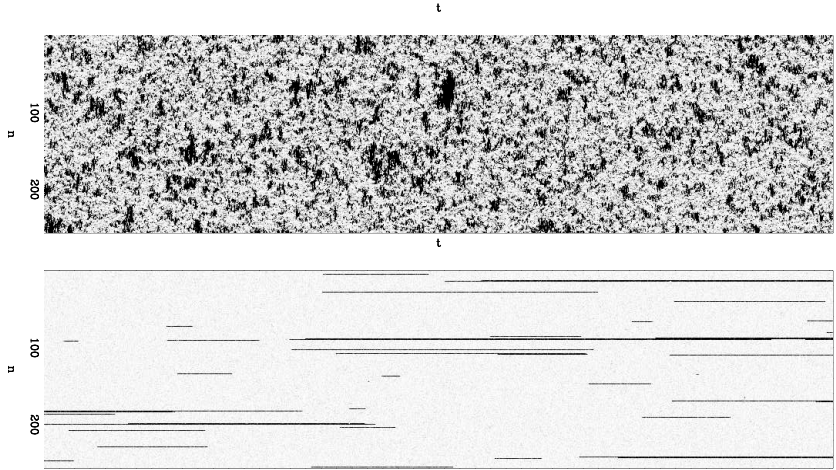

  \centering
  \includegraphics[angle=270,width=11cm]{%
greyy1123.eps}

  \includegraphics[angle=270,width=11cm]{%
greyy1126.eps}
  \caption{Comparison of the dynamics of the model with the Morse
    potential $V(y)$ (top figure) and with the modified potential with
    a hump $V_h(y)$ (bottom figure)
obtained from a numerical simulation of the models
in contact with a
thermal bath at $270\;$K. The stretching of the base pairs is shown by
a grey scale going from white for a closed pair to black for a fully open
pair $y \ge 2.0\;$\AA. 
The vertical axis extends along the DNA chain, which has 256
base pairs in these calculations, with periodic boundary
conditions. The horizontal axis corresponds to
time.
The parameters used for this calculation are 
$K = 0.01\;$eV \AA$^{-2}$, $\rho = 3.0$,
$\delta = 0.8\;$\AA$^{-1}$,
$D = 0.05254\;$eV,
$\alpha = 4.0\;$\AA$^{-1}$ for the Morse potential (top figure) and
$\alpha = \beta =
    4.0\;$\AA$^{-1}$, $E = 4.0\;$\AA$^{-1}$ and  $D = 0.0857\;$eV for
    the potential $V_h(y)$ (bottom figure).
The total time shown in these figures is $2\;10^{-8}\;$s.
}
  \label{fig:greynewmodel}
\end{figure}

Ideas to solve this discrepancy come from all-atom molecular dynamics (MD)
simulations \cite{GIUDICE}. As opening is a rare event, which may
extend over hundreds of nanoseconds, MD simulations which have to
study a very large number of degrees of freedom, including those of
the solvent, cannot study the lifetime of the open state. But these
simulations can be biased to observe the free energy pathway
associated with opening. This is done by adding a geometrical restraint
imposed by a bias potential that imposes a given opening. Then the
fluctuations of the DNA structure are recorded for all intermediate
positions. Their probability distribution, corrected from the effect
of the bias, gives the free energy as a function of the opening
\cite{GIUDICE}. The results are highly sensitive to the details of the
solvent and counter ions dynamics, but they nevertheless show that the
free energy of the open state may have a shallow minimum. This is
consistent with a dynamics in which the base pair may stay open for a
long time rather than vibrating as a breather. Moreover the
simulations show that open bases can fluctuate a lot. Their motions
include rotations, which may hinder the reclosing of the pairs, 
inducing an energetic barrier for closing. In a mesoscopic model which
only describes a subset of the degrees of freedom of the system, the
potentials are actually free energies that take into account all the
degrees of freedom which are not included in the model. Therefore it
is natural to replace the simple Morse potential by a more elaborate
function which describes the effects discussed above, and particularly
the barrier for reclosing. To allow analytical calculations we have
chosen the following expression \cite{PCJ}
\begin{equation}
  \label{eq:potj}
V_h(y) = \left\{
  \begin{array}{ll}
    A \big[ e^{-\alpha y} - 1 \big]^2 & \mbox{~~if~~ $y < 0$},\\
    a y^2 + b y^3 + c y^4 & \mbox{~~if~~ $0 \le y \le 1$},\\
    D + E e^{-\beta y} \Big(y + \frac{1}{\beta}\Big) & \mbox{~~if~~
      $y>1$} \; ,
  \end{array} \right.
\end{equation}
which is determined when the parameters $D$, $E$, $\alpha$, $\beta$
are selected. The other parameters are calculated to ensure the
continuity of the potential and of its first and second derivatives.
For $y<0$ this potential is identical to a Morse potential, while for
large values of $y$ it decreases towards $D$ after a hump which
describes the barrier for reclosing. The polynomial form for the
intermediate range of $y$ provides a smooth matching between the two
domains. This new model brings the qualitative improvement needed to reconcile the theory and the experiments because the dynamics of the model in contact with a thermal bath is drastically modified, as shown in Fig.~\ref{fig:greynewmodel}.
For the Morse potential, with a set of parameters realistic for DNA, the average life time of an open base pair is only $0.08\;$ns, while it increases to $7\;$ns when we use the new potential, bringing it in the range
measured in experiments \cite{GUERON1990}. Moreover the open states
concern only one or two consecutive bases, in agreement with the
observations \cite{Leroy} while the Morse potential was giving larger
bubbles, even well below the melting transition. The average lifetime of a closed base pair is found to be $0.4\;$ns for the Morse potential whereas we get $0.4\;\mu$s when we use the new model. This value is still well below the experimental estimates of a few ms \cite{Leroy}, but the new potential nevertheless improves the results by three orders of magnitude.

\bigskip
The analysis of the timescales in the dynamics of the model provides a second example of the interest of the study of a system like DNA, which can be probed by a large variety of experimental methods. It illustrates how the comparison between theory and experiment has been fruitful to reach a model which is much more satisfactory than the original nonlinear lattice model of DNA. But this does not mean that all difficulties to describe DNA theoretically are solved. There are still open problems, currently under investigation but not yet satisfactorily solved.

\section{Predicting the melting of short sequences: a challenge}
\label{sec:sequence}

The most interesting problem for DNA is the study of the role of the sequence on its properties, i.e.\ the influence of the genetic code. It is a challenge because it combines nonlinearity and heterogeneities. 
DNA contains 4 types of bases, labelled A,T,G,C, which can form only two types of pairs AT and GC. However, because the stacking depends on the relative position of the bases, AT followed by AT is not equivalent to 
AT followed by TA, so that there are 16 possible stacking interactions between adjacent pairs. Moreover AT pairs are linked by 2 hydrogen bonds while GC pairs are linked by three bonds. Thus AT pairs are easier to break and this difference can be reflected in the model by two types of potentials $V(y)$.

The thermal denaturation of a natural DNA segment is a complex process because some regions of the molecule open at lower temperature than others, giving an experimental denaturation pattern which is specific of the sequence. Understanding the origin of this pattern with a model is still an open problem. For very long sequences, with tens of thousands of base pairs, experiments do not resolve a fine structure and detect the opening of regions that extend over hundreds of base pairs. In this case Ising models, with an appropriate set of parameters, can be very successful \cite{WADA}. On the contrary, for short sequences with a few tens of base pairs, the challenge persists. This scale is important because it is the scale of many biological phenomena which involve the binding of proteins to DNA. For instance transcription starting domains are regions where the reading of genes starts. They have a size of a few tens of base pairs and, as transcription requires a local opening of DNA, a correct understanding of the mechanism of this opening could help in the detection of these biologically important regions. This has not escaped to the attention of physicists and the simple model described above has been used to try to detect transcription start sites. The optimistic conclusions of the first attempt \cite{CHOI} were however softened by further studies using more accurate statistical methods \cite{VANERP}.

Experiments on the melting of short DNA segments are now becoming very accurate, and, selecting special sequences, it is possible to determine whether, in the intermediate stages, all molecules are partly open or whether some are closed while others are fully open \cite{Montrichok,ZENG}. It is also possible to make a mapping of the local structure and fluctuation of all GC pairs in a DNA segment versus temperature \cite{Angelov} with a method that uses a mixture of physical and biological techniques: a short laser pulse ionises DNA by bi-photonic excitation and the degradation products, which depend on the state of the molecule when it was hit by the laser, are analysed by standard biological methods. Among all the recent experimental results on sequence-dependent melting of DNA, one exemplifies the difficulty of the analysis.
\begin{figure}[h]
\begin{center}
\includegraphics[width=\textwidth]{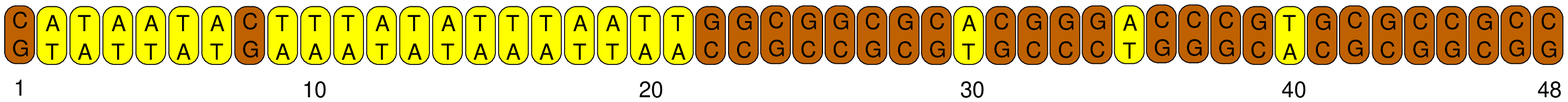}
\caption{Sequence studied in Ref.~\cite{Montrichok}. It shows a two-step melting. More than 75\% of the sequence opens first, and the remaining part opens about 10K above.}
\label{fig:Montrichok48}
\end{center}
\end{figure}
Its sequence, shown in Fig.~\ref{fig:Montrichok48}, contains an AT-rich domain that covers about half of the sequence while the second half is dominated by stronger GC base pairs.
One would expect it to denaturate in two steps with the AT-rich region opening first, and then the second half opening at a higher temperature. But it is not so. The denaturation does take place in two steps, but approximately 75\% of the molecule opens in the first step. This implies that the opening occuring in the lowest temperature range extends well inside the supposedly strong GC-rich region. Standard programs used by biologists to predict denaturation profiles of DNA sequences, based on the Ising model \cite{Dinamelt}, fail to predict the correct denaturation curve.

To apply the nonlinear model to this question, the sequence must be introduced in the Hamiltonian. The simplest approach keeps a uniform stacking interaction, and distinguishes the AT and GC base pairs by the on-site potential $V(y)$ \cite{Campa}. It can give satisfactory results in some cases, but fails, like the Ising models, for the example of Fig.~\ref{fig:Montrichok48}. Taking into account stacking inhomogeneities introduces a large number of additional parameters, so that one could think that it is sufficient to fit any experimental result.
This approach has been used for the Ising model, and as expected, it can provide a rather good agreement with experiments {\it if the parameters are adjusted for each sequence of interest} \cite{Ivanov}. This is not very satisfactory. One would like to find a DNA model that can be applied to any sequence and predict its melting curve.

This is still an open question but, as for the other features of the model discussed above, hints are given by the experiments. Using bi-photonic ionisation of the guanines, we have measured the fluctuations of synthetic DNA sequences containing a large AT-rich domain \cite{Cuesta}. The results show that the thermal fluctuations of this weaker region influence the structure and fluctuations of DNA about 10 base pairs away.
This indicates that {\em non-local effects are important in DNA.} This explains why the theoretical description of the melting of
short sequences is such a challenge. Models based on local effects can only give a rough picture of the reality. In many cases this is sufficient, but in specific cases such as the sequence of Fig.~\ref{fig:Montrichok48}, these models fail badly.

\section{Other open problems}
\label{sec:open}

The difficulty that we mentioned for short sequences could well be related to the three-dimensional structure of DNA. In the models at the scale of a base pair that we described in the previous sections, this structure is totally ignored. It would be pointless to try to go to the other extreme and introduce all atomic coordinates in a model. This is done in molecular dynamics simulations, which must also include the water molecules of the solvent and counter ions to give correct results, but it leads to extremely heavy calculations. A one-month simulation on a super-computer is something common when one intends to reach this level of details in the description. The interesting problem for nonlinear science is to determine what is essential, and restrict modelling to these aspects only. For DNA there is something that comes to mind immediately, it is the helicoidal structure. In the three-dimensional space, for a helix opening and twist are coupled. This can lead to interesting theoretical problems which have been considered for DNA \cite{BarbiCocco,BarbiLepri} in the particular case when the axis of the helix stays straight. 
This hypothesis significantly simplifies the model, but it is not realistic. For instance it is likely that the non-local effect discussed above for the melting of some sequences is related to a coupling between torsion, twist and opening, which is affected by the opening of the AT-rich region.

\medskip
Even if one stays at the level the simple model with one degree of freedom per base pair, there is another open problem. It concerns the temperature range over which the transition occurs in short DNA molecules. In Sec.~\ref{sec:model} we showed how the introduction of a nonlinear stacking interaction can sharpen the denaturation transition. But this result is obtained by a statistical physics analysis which is valid for an infinite lattice. When the calculation is performed for a short DNA chain, the transition broadens by finite size effects. This is expected but the problem is that this broadening is too large compared to experiments. This has a consequence on the calculated melting curves of short sequences because, for a molecule made of two segments that should show two distinct transitions, the model gives instead two broad overlapping melting ranges. As in the infinite case, this has to do with the entropy of the open state, which is not properly described with a single degree of freedom per base pair. In the infinite case the nonlinear stacking can bring a sufficient entropy because the open regions are very large, allowing extensive fluctuations. In a short DNA molecule the simple model shows its shortcoming, and a more complete description of the open state would be necessary. The method that we introduced for DNA hairpins, which uses a three-dimensional polymer model to describe the open regions of the molecule \cite{Errami}, might provide a solution to this problem, but it is hard to combine this approach with a model for the melting of the double helix.

\medskip
Another open problem for the mesoscopic modelling of biomolecules is the definition of a correct thermostat.  In numerical simulations one generally uses Langevin or Nose thermostats, which are valid when all the degrees of freedom of the system of interest are described. In the case of DNA, it is easy to realise that, when a base pair is closed it is only weakly coupled to the solvent because the two bases are hidden inside the double helix. On the contrary, when the base pair is open, the bases are in close contact with the solvent. In a model where the opening is only described by a single variable $y$, its coupling with the thermostat should depend on the value of the variable. Some studies have been initiated in this direction, but the question is still deserving attention \cite{Samoletov}.

\medskip
Finally we would like to point out that modelling DNA raises new interesting questions for nonlinear science itself. The potential (\ref{eq:potj}) that we introduced to get the correct time scales for the DNA fluctuations in the model sustains breather-like localised excitations which exhibit a non-standard amplitude-energy curve \cite{PCJ}. These solutions do not oscillate around the
ground state of the system, but around an exponentially localised equilibrium configuration $y_n (K)$ corresponding (for a harmonic coupling with $K\approx 0$) to the opening of a unique base pair. Interestingly, if $K \rightarrow 0$ the stretching of the open base pair at $n=n_0$ diverges logarithmically (${y}_{n_0} (K)\sim -\frac{1}{\beta}\, \ln{(K)}$) while ${y}_n (K) \rightarrow 0$ for the other unexcited bases at $n\neq n_0$. This yields interesting mathematical problems, namely to analyze the movability of such localized states and to prove the existence of discrete breathers having similar properties. This point would require an extension of MacKay and Aubry's theory \cite{MACKAY94} allowing critical points at infinity.

\end{document}